\documentclass[12pt]{article}
\usepackage{amsmath,amssymb,epsfig}
\usepackage{graphicx}

\newcommand{\be}{\begin{equation}}
\newcommand{\ee}{\end{equation}}
\newcommand{\ea}{\end{array}}
\newcommand{\beqa}{\begin{eqnarray}}
\newcommand{\eeqa}{\end{eqnarray}}

\def\CP2{{\mathbb C}P^2}

\def\CDalign#1{\bgroup\vcenter\bgroup\tabskip 2pt 
       \baselineskip 14pt \lineskip 3pt \lineskiplimit 3pt
       \halign\bgroup &\hfill$##$\hfill\crcr
       #1\crcr\egroup\egroup\egroup} 

\newcommand{\gapproxeq}{\lower .7ex\hbox{$\;\stackrel{\textstyle
>}{\sim}\;$}}
\newcommand{\lapproxeq}{\lower .7ex\hbox{$\;\stackrel{\textstyle
<}{\sim}\;$}}

\newcounter{appendice}

\def\thebibliography#1{{\bf REFERENCES\markboth
 {REFERENCES}{REFERENCES}}\list
 {[\arabic{enumi}]}{\settowidth\labelwidth{[#1]}\leftmargin\labelwidth
 \advance\leftmargin\labelsep
 \usecounter{enumi}}
 \def\newblock{\hskip .11em plus .33em minus -.07em}
 \sloppy
 \sfcode`\.=1000\relax}

\begin{document}
\begin{titlepage}
\title{An equation of state for purely kinetic k-essence inspired by
cosmic topological defects
\author{
Rub\'en Cordero \footnote{Email: cordero@esfm.ipn.mx}, \,
 Eduardo L. Gonz\'alez \footnote{Email:
egonzalez@esfm.ipn.mx}\,\, and Alfonso Queijeiro \footnote{Email:
aquei@esfm.ipn..mx}  \\
{\small\it Departamento de F\'\i sica, Escuela Superior de F\'\i sica
y Matem\'aticas del Instituto Polit\'ecnico Nacional}\\
{\small\it Unidad Adolfo L\'opez Mateos, Edificio 9, 07738 Ciudad de M\'exico, MEXICO}\\
}}
\maketitle
\begin{abstract}
We investigate the physical properties of a purely kinetic k-essence model with an
equation of state motivated in superconducting membranes. We compute the equation of state parameter
$w$ and discuss its physical evolution via a nonlinear equation of state. Using the adiabatic speed of
sound and energy density, we restrict the range of parameters of the model in order to have an acceptable
physical behavior. Furthermore, we analyze the evolution of the luminosity distance $d_{L}$ with redshift $z$
by comparing (normalizing) it with the $\Lambda$CDM model.
Since the equation of state parameter is $z$-dependent the evolution of the luminosity distance is also
analyzed using the Alcock-Paczy\'{n}ski test.
\end{abstract}
\end{titlepage}

\section{Introduction}

The discovery of the accelerating expansion of the universe \cite{Riess,Perlmutter} is one of the most intriguing
and puzzling questions in cosmology today \cite{Sahni:2004ai,Alam:2004jy,Sahni:2006pa,Feng:2004ad,Durrer:2007re}.
There are several attempts to address this question and they include: the cosmological constant, quintessence
field \cite{Wette,Peebles,Copeland,Linder:2007wa}, brane cosmology scenario \cite{Deffayet,SS} and k-essence
models \cite{InfADM,Armendariz-Mukhanov-Steinhardt,AMS,Melchiorri,Chiba1,Chiba2,Chimento1,Chimento2004}. Among the former
proposals the cosmological constant is the simplest choice although it needs a fine tuning value. There are different
approaches that try to model the late-time accelerated expansion of the universe including modified gravity at
large distances via extradimensions or including a dark energy component that allows accelerating universes by means
of a kind of energy that does not fulfill the strong energy condition \cite{Amendola-Tsujikawa}

Scalar fields models with non-canonical kinetic terms have been proposed as an alternative to describe the
dark energy component of the universe. The k-essence type lagrangians were introduced in several contexts,
for example, as a possible model for inflation \cite{InfADM,Jaume}. Later on k-essence models were analyzed
as an alternative to describe the characteristics of dark energy and as a possible mechanism of unifying dark
energy and dark matter \cite{Scherrer:2004au}. Purely kinetic k-essence models \cite{Linder1,Gao,holographic,crossing}
are, in some sense, as simple as quintessence because they depend only on a single function $F$ by means of the
expression for the density Lagrangian ${\cal{L}} = F(X)$, where the kinetic term is $X$.
It is very interesting to note that the form of the purely k-essence Lagrangian appear in superconducting membranes
where it is constructed by means of a spontaneously symmetry breaking of an underlying field theory
\cite{Witten,Alexbook,CarterRev,Peter2}.
The scalar field is related to the $U(1)$ symmetry breaking corresponding to the electromagnetic field
and it is responsible for the existence of permanent currents. This kind of fields have been studied for superconducting
strings and walls \cite{supersonic,Lazarides:1985my,Peter:1995ks}. Following the idea of Rubakov and Shaposhnikov
\cite{Rubakov:1983bb}, who postulated that our universe can be viewed as a topological defect, we want to introduce
the k-essence field in the form of a superconducting membrane field. In this proposal we can start with a 5-dimensional
field theory and follow a similar pattern for the formation of a superconducting membrane. We postulate that our universe
can be viewed as a topological defect and the k-essence field emerges in a similar fashion as a scalar field from a
spontaneously symmetry breaking of the underlying field theory. It is worth to mention that several equations of
state have been studied including $\rho = p + constant$ \cite{Witten} and the so called permanently transonic or
Nielsen model for superconducting strings and membranes of the form $\rho p =-constant$ \cite{CarterRev,Nielsen,Cordero:2000hx}.
It is very important to remark that this equation of state corresponds to a k-essence Chaplygin gas model
\cite{Gorini-Kamenshchik-Moschella-Pasquier,Kamenshchik-Moschella-Pasquier,Bento-Bertolani-Sen}.

It has been argued, in the context of superconducting extended objects \cite{supersonic}, that a polynomial type
equation of state of the form\footnote{Usually the equation of state is written in the form $p=p(\rho)$.}
$\rho = d + b p + cp^2$ (where $d$,$b$, and $c$ are constants related to the mass of the charge carriers
and the Kibble mass) is a better equation of state in order to describe some properties of this kind of objects.
As a first approach we would like to study this kind of equation of state. However, it remains to develop the complete
construction of a 5-dimensional field theory and to obtain the possible modifications that can arise from
higher dimensions. In addition, we can consider this kind of equation of state as an expansion in Taylor
series of $\rho(p)$ around an specific value of the pressure. Inspired by these ideas we study
in this work a generalized polynomial equation of state in the context of k-essence models giving an additional
physical meaning for the scalar field. In this paper we also analyze the adiabatic speed of sound and restrict the
allowed values for the parameters included in the equation of state in order to satisfy stability and positive
energy and, besides, we find the region for subluminal speed of sound. Also we study the evolution of the luminosity distance
$d_{L}$ with redshift $z$ comparing (normalizing) it with the $\Lambda$CDM model and also applying the Alcock-Paczy\'{n}ski test.

This paper is organized as follows. In Section 2 we present the essential characteristics of purely kinetic k-essence fields.
In Section 3 we analyze the polynomial type equation of state and establish the physical allowed region in the space of parameters.
In Section 4 we study some cosmological properties of the parameters of the model.
Finally, in the last section we give our conclusions.

\section{Purely kinetic k-essence}
The k-essence model was proposed originally in \cite{InfADM}, \cite{Armendariz-Mukhanov-Steinhardt}, \cite{AMS} as a
model to describe the accelerated expansion of the universe within the context of inflation. The
k-essence field became later a candidate to dark energy \cite{Melchiorri,Chiba1,Chiba2},
and as such it is defined as a scalar field $\phi$ with non-linear kinetic energy terms in $X$ according to the action
\begin{equation}
S = \int d^4 x \sqrt{-g}F(\phi, X),
\end{equation}
where $X = \frac{1}{2}\partial_\mu \phi \partial^\mu\phi$. Furthermore, in this work we focus our attention on
purely kinetic k-essence models described by the action
\begin{equation}
S = \int d^4 x \sqrt{-g}F(X).
\end{equation}
The equation of motion for the field $\phi$ is obtained in the usual way from the last action and is written in the form
\begin{equation}
\nabla_\mu J^\mu = 0,
\end{equation}
where the conserved current is \cite{CarterRev,Cordero:1999xd}
\begin{equation}
J^\mu =  F_X g^{\mu \nu}\partial_\nu \phi,
\end{equation}
where $F_{X} \equiv \frac{dF}{dX}$. The conservation of the current is equivalent
to the conservation of the energy-momentum tensor \cite{superhybrid} of the k-essence field
\begin{equation}
T^{\mu \nu} = F_X \partial^\mu \phi\partial^\nu \phi - g^{\mu \nu}F.
\end{equation}
We consider the Friedmann-Robertson-Walker (FRW) metric
\begin{equation}
ds^2 = dt^2 - a^2(t)d\overrightarrow{x}^2,
\end{equation}
and the field $\phi$ to be smooth, thus $X=\frac{1}{2}\dot{\phi}^2$, i.e. $X$ is positive.
We have a comoving perfect fluid energy-momentum tensor where the k-essence energy density $\rho$ and the
pressure $p$ are given by
\begin{equation}
\rho = 2XF_X - F, \label{energy}
\end{equation}
and
\begin{equation}
p = F(X). \label{pressure}
\end{equation}
The conserved current gives as a result the following relation between the $X$ parameter and the scale factor $a$
\begin{equation}
XF_X ^2 = k_{0}a^{-6}, \label{scalefactor}
\end{equation}
where $k_{0}$ is a constant of integration. This solution was derived for the first time by Chimento in \cite{Chimento2004}.
Given a form of $F(X)$, Eq. (\ref{scalefactor}) gives a relation between $X$ and $a$, then the evolution of the equation of
state parameter $w$ and the sound speed $c^{2}_{s}$ as a function of the scale factor $a$.
The equation of state parameter $w$ in k-essence models has the form
\begin{equation}
w = \frac{p}{\rho}= \frac{p}{2Xp_X - p}. \label{EofS}
\end{equation}
and the adiabatic speed of sound is given by
\begin{equation}
c_s ^2 \equiv \frac{\partial p/\partial X}{\partial \rho/\partial X} = \frac{F_X}{2XF_{XX} + F_X}
= \frac{F^{2}X}{(XF^{2}_{X})_{X}} \label{speed}
\end{equation}
with $F_{XX} \equiv d^{2}F/dX^{2}$. In order to satisfy the stability condition $c_s ^2 \geq 0$,
the last equation imposes the relation $\frac{da}{dX} < 0$, after using eq. (\ref{scalefactor}).

\section{The model}

We consider a generic equation of state of the form
\begin{equation}
\rho = d + bp + cp^2, \label{polynomial}
\end{equation}
where $d$, $b$, and $c$ are constants and their range of validity will be determined by means of physical requirements.
The form of the last equation of state is inspired by the study of superconducting membranes (see \cite{CarterRev}).
It is interesting to note here that an analagous equation of the form $p = \lambda + \delta \rho + \kappa \rho^2$ has been
introduced in \cite{Odintsov1}, furthermore studied in \cite{Odintsov2,Odintsov3,Odintsov4}, and subsequently generalized in 
\cite{Odintsov5}. A general review of different fluids models with such a generalized equation of state is given in \cite{Odintsov6}.
The relevant physical quantities $p, \rho, w$, and $c_s ^2$ can all be computed using Eqs. (\ref{energy}), (\ref{pressure})
together with Eqs. (\ref{EofS}), (\ref{speed}). In order to obtain the form of $F(x)$ we use equations
(\ref{energy}) and (\ref{polynomial}) so that the equation we have to integrate is the following
\begin{equation}
2X \frac{dp}{dX} = d + (b + 1)p + cp^{2}. \label{integrate}
\end{equation}
If $\Delta \equiv 4cd -(b + 1)^{2}$ the integration of equation (\ref{integrate}) leads to
three different solutions depending on $\Delta < 0$, $\Delta > 0$, or $\Delta = 0$.

\subsection{The case $\Delta<0$}
For this case it is found the following expression for $p$
\begin{equation}
F(X) = p = \frac{1}{2c}\left(\frac{\sqrt{-\Delta} - b - 1 + \tilde{c}(\sqrt{-\Delta} + b + 1)X^{\alpha}}
{1 - \tilde{c}X^{\alpha}}\right) \label{p1}
\end{equation}
where $\tilde{c}$ is an integration constant, and we choose it to be $\tilde{c} = 1$.
Defining $\alpha$ through the expression $2\alpha = \sqrt{-\Delta}$, and also $\beta = \alpha/c$ and $\gamma = (b+1)/2c$,
Eq. (\ref{p1}) finally becomes
\begin{equation}
p = -\beta -\gamma + \frac{2\beta}{1 - X^{\alpha}}.
\end{equation}
In the particular case where $\alpha = 1$ our model coincides with one of the models introduced
by Carter and Peter in \cite{supersonic} where they provide a realistic representation of the macroscopic dynamical
behavior of Witten-type (superconducting) vortex defects \cite{Witten}.

The energy density could be written in the form
\begin{equation}
\rho = \frac{1}{2c}\left(\frac{(2\alpha + b + 1)X^{2\alpha}
+ (8\alpha^2 - 2b - 2)X^\alpha + b + 1 - 2\alpha}{(1 - X^\alpha)^2}\right). \label{k-essence-rho}
\end{equation}

The equation of state parameter $w$ is found to be
\begin{equation}
w = \frac{-(2\alpha + b + 1)X^{2\alpha} + 2(b + 1)X^\alpha + 2\alpha - b - 1}{(2\alpha + b + 1)X^{2\alpha}
+ (8\alpha^2 - 2b - 2)X^\alpha + b + 1 - 2\alpha}. \label{stateparameter}
\end{equation}
It is worth to mention that the kinetic k-essence Lagrangian has the scaling property ${\cal{L}} \rightarrow \kappa {\cal{L}}$,
that leaves the equation of state parameter and the speed of sound unchanged, and as a consequence it gives only two relevant
parameters (for example, $b$ and $cd$).

The expression for the scale factor $a$ as a function of the parameter $X$ is
found from Eq. (\ref{scalefactor}) and is given by
\begin{equation}
a = \frac{k^{1/6}c^{1/3}}{(2 \alpha^{2})^{1/3}} \cdot \frac{(1 - X^\alpha)^{2/3}}{X^{(2\alpha - 1)/6}}. \label{sf}
\end{equation}
Taking $\alpha = 1$ and defining $r \equiv \frac{k^{1/6}c^{1/3}}{(2 \alpha^{2})^{1/3}}$ we have finally
\begin{equation}
r^{-3/2} \cdot a^{3/2} X^{1/4} + X = 1, \label{quartic-equation}
\end{equation}
Our aim is to solve $X$ as a function of the scale factor $a$ and $r$. Eq. (\ref{quartic-equation}) admits
four solutions, two of them are real, and from these we consider the one physically acceptable.

\begin{figure}
\begin{center}
\includegraphics[scale=1]{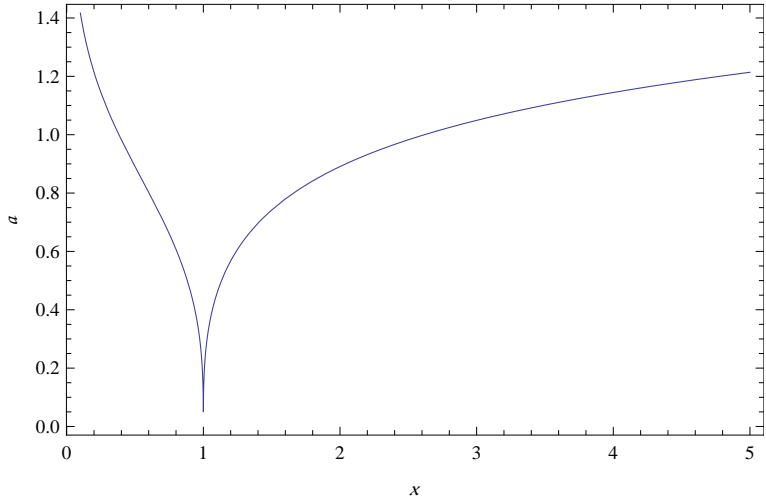}
\caption[Short caption for figure 1]{\label{fig:scalefactor} {Scale factor as a function of $X$ for $\alpha = 1$.
For simplicity, we have chosen the numerical pre-factor in eq. (\ref{sf}) equal to one.}}
\end{center}
\end{figure}

In figure \ref{fig:scalefactor} we present the relation between the scale factor $a$ and the parameter $X$ for $\alpha = 1$.
In order to satisfy the stability condition $da/dX < 0$ the values of $X$ must be restricted to $0 < X < 1$.
It is observed the following correspondence: when $X \rightarrow 1$ then $a \rightarrow 0$ and when $X \rightarrow 0$
then $a \rightarrow \infty$ if $\alpha > \frac{1}{2}$.

\begin{figure}
\begin{center}
\includegraphics[scale=0.8]{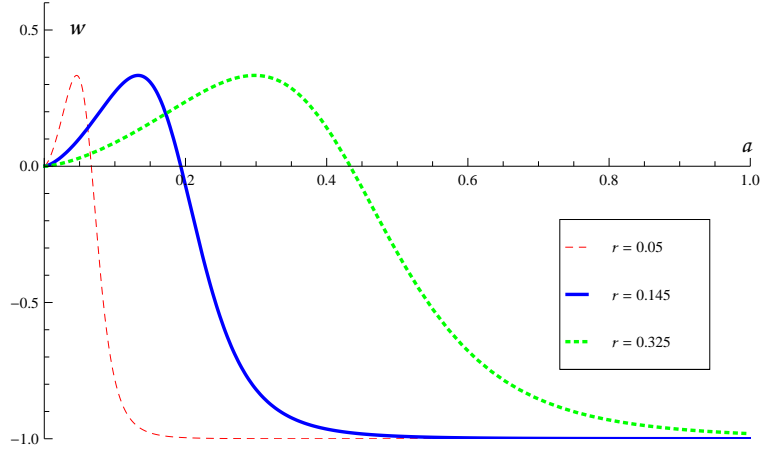}
\caption[Short caption for figure 2]{\label{fig:parametro-estado(alpha1b3)}
{\scriptsize Equation of state parameter $w$ as a function of $a$ for different values of $r$.
All curves correspond to $\alpha=1$ and $b=3$.}}
\end{center}
\end{figure}

\begin{figure}
\begin{center}
\includegraphics[scale=0.8]{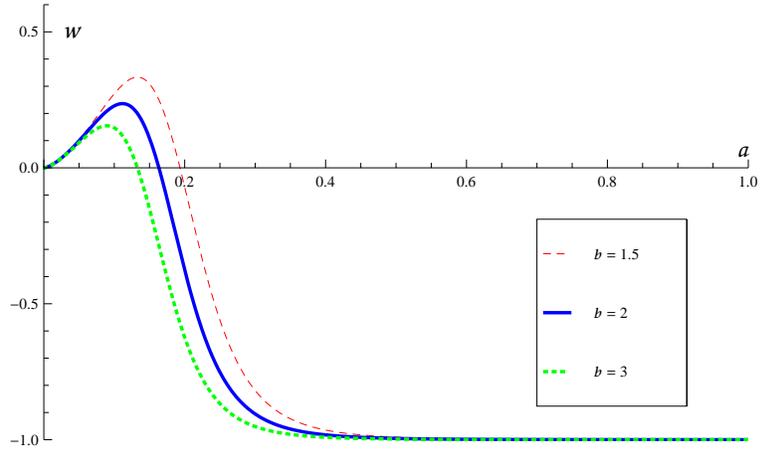}
\caption[Short caption for figure 3]{\label{fig:parameteralpha1r145}
{\scriptsize Equation of state parameter $w$ as a function of the scale factor $a$ for different values of $b$
All curves correspond to $\alpha=1$ and $r=0.145$.}}
\end{center}
\end{figure}

\begin{figure}[tbp]
\begin{center}
\includegraphics[scale=0.9]{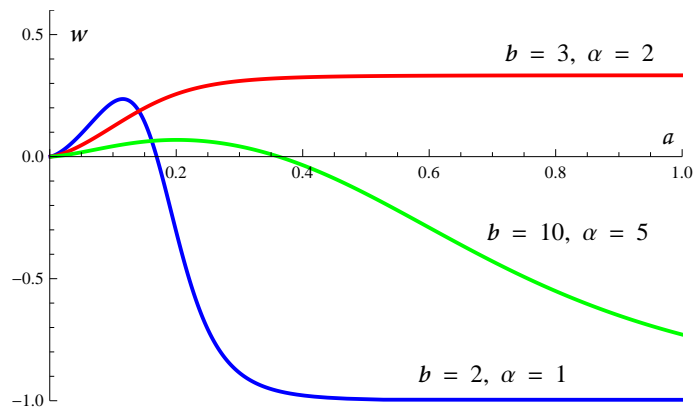}
\caption[Short caption for figure 4]{\label{fig:varios-alpha-b}
{\scriptsize Equation of state parameter $w$ as a function of the scale factor $a$. All curves correspond
to $r=0.15$}}
\end{center}
\end{figure}

In figure \ref{fig:parametro-estado(alpha1b3)} we present the evolution of the equation of state parameter $w$
as a function of the scale factor $a$ for different values of $r$ but with a same value for the
parameter $b$ and for $\alpha = 1$. It is observed that at the beginning of its evolution the universe has a dust-type
dominant component, furthermore it evolves into radiation ($w=1/3$), then again has a dust-type (matter) component, and finally
it tends asymptotically to a cosmological constant ($w=-1$). On the other hand, in figure \ref{fig:parameteralpha1r145} it is
represented the evolution of $w$ as a function of the scale factor $a$ for different values of $b$ with a same value
for $r$. In figure \ref{fig:varios-alpha-b} it is shown the evolution of the state parameter $w$ as a function of $a$ for
various values for $\alpha$ and the parameter $b$.

\begin{figure}
\begin{center}
\includegraphics[scale=0.9]{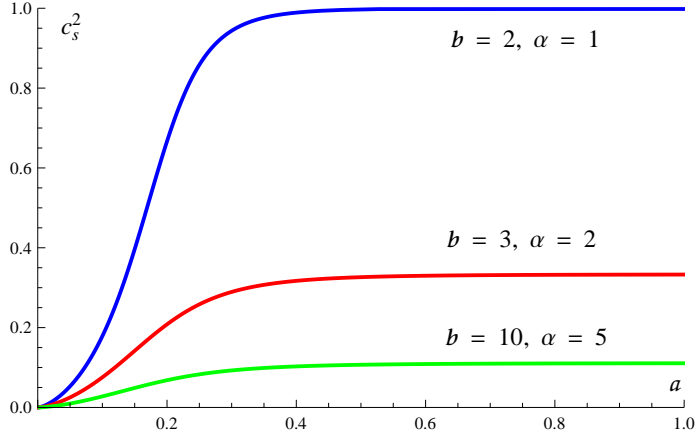}
\caption[Short caption for figure 5]{\label{fig:velocitya1b2}
{\scriptsize Adiabatic speed of sound $c^{2}_{s}$ as a function of the scale factor $a$ for $r=0.15$.}}
\end{center}
\end{figure}

From definition (\ref{speed}) the expression for the adiabatic sound speed is found to have the form
\begin{equation}
c_s ^2 = \frac{1 - X^\alpha}{2\alpha - 1 + (2\alpha + 1)X^\alpha}.
\end{equation}
We would like to satisfy the physical requirement
\begin{equation}
0\leq c_s ^2 \leq 1.
\end{equation}
The stability condition $c^{2}_{s}\geq0$ implies $\frac{1}{2} \leq \alpha$ and the positive energy requirement
gives $b + 1\geq 2\alpha$. The subluminal sound speed condition produces the inequality $1\leq \alpha$.
In fact, there is an open debate about the possible existence of acceptable physical systems with superluminal
speed of sound \cite{superluminal1,Bruneton:2006gf,Kang:2007vs,Bonvin:2006vc,Gorini:2007ta}.
The positive energy condition is satisfied in this model if $dc > 0$.

The evolution of the sound speed is represented in figure \ref{fig:velocitya1b2} and we can
appreciate that it tends to a constant. In fact, when $2\alpha \neq b + 1$, the speed of sound takes the value
$c_s ^2 = \frac{1}{2\alpha -1}$ when the scale factor is very large. When the condition $2\alpha = b + 1$ is
fulfilled the evolution of the equation of state parameter does not exhibit a cosmological constant behavior,
but for large values of the scale factor $a$ it reaches the value $w = \frac{1}{b}$.

\subsection{The case $\Delta=0$}
For this case the expression for pressure is given by
\begin{equation}
p = \frac{1}{2c} \left(- b - 1 - \frac{4}{\ln X} \right).
\end{equation}
The energy density is
\begin{equation}
\rho = \frac{1}{2c}\left( b+1 + \frac{2}{\ln X} + \frac{8}{(\ln X)^2} \right).
\end{equation}
The equation of state parameter is written in the form
\begin{equation}
w = \frac{\left(- b - 1 - \frac{4}{\ln X} \right)}{\left( b+1 + \frac{2}{\ln X} + \frac{8}{(\ln X)^2}\right)}.
\end{equation}
The relation between the scale factor $a$ and the parameter $X$ is given by
\begin{equation}
a = \left(\frac{kc^2}{4}\right)^{1/6} X^{1/6} (\ln X)^{2/3}.
\end{equation}
The stability condition is not satisfied in this case because there is no region where $da/dX<0$.

\subsection{The case $\Delta>0$}
The expression for the pressure becomes
\begin{equation}
p = \frac{1}{2c}\left(- b - 1 + 2\beta \tan\left(\frac{\beta \ln X}{2}\right)\right),
\end{equation}
where $2\beta = \sqrt{\Delta}$. The energy density is given by
\begin{equation}
\rho = \frac{1}{2c}\left(\beta \sec^2 \left( \frac{\beta \ln X}{2}\right)
- 2\beta \tan\left(\frac{\beta \ln X}{2} \right) + b + 1 \right).
\end{equation}
The equation for the state parameter $w$ is
\begin{equation}
w = \frac{- b - 1 + 2\beta \tan\left(\frac{\beta \ln X}{2}\right) }{ \beta \sec^2 \left( \frac{\beta \ln X}{2}\right)
- 2\beta \tan\left( \frac{\beta \ln X}{2} \right) + b + 1}.
\end{equation}
For this case the expression of the scale factor $a$ as a function of the parameter $X$ is found to be
\begin{equation}
a = \left(\frac{4kc^2}{\beta^4 }\right)^{1/6} X^{1/6} \cos^{2/3}\left(\frac{\beta}{2}\ln X\right). \label{oscillatory}
\end{equation}
Because of the oscillatory behavior of the scale factor with respect to $X$ in Eq. (\ref{oscillatory}), this solution
is not adequate to describe the evolution of the universe, especially when the scale factor becomes large.

\begin{figure}
\begin{center}
\includegraphics[scale=0.7]{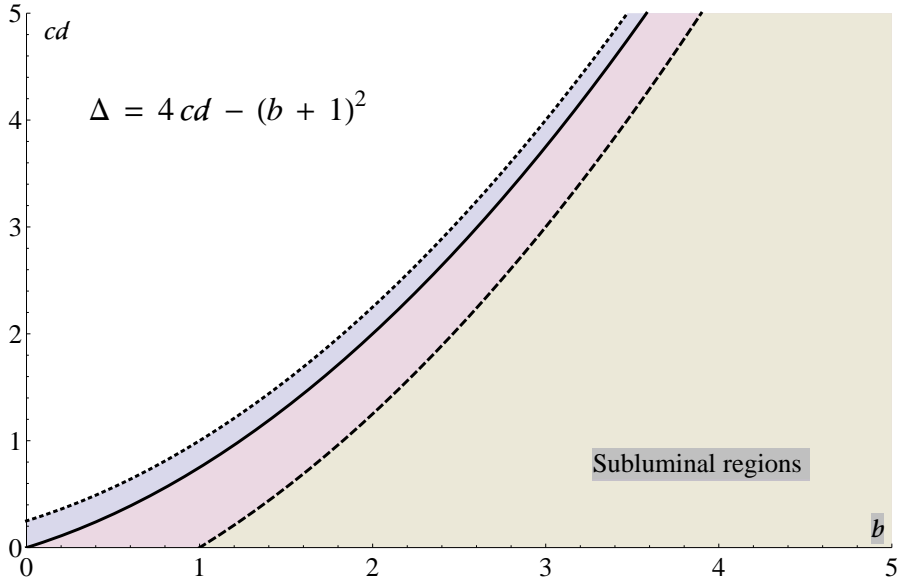}
\caption[Short caption for figure 6]{\label{fig:allowed}
{Allowed physical regions for the parameters $cd$ and $b$. The dotted line corresponds
to points for which $\Delta = 0$}, the full line for points where $c^{2}_{s} = \infty$, and the
dashed line for points where $c^{2}_{s} = 1$.}
\end{center}
\end{figure}

Finally, in figure \ref{fig:allowed} we present the physical allowed regions for the parameters $cd$ and $b$ of the model.
It is possible to include higher exponents in the polynomial equation of state, but there is not a general analytical
expression for the pressure and the energy.

\section{Cosmological analysis of the parameters}
\begin{figure}
\begin{center}
\includegraphics[scale=0.7]{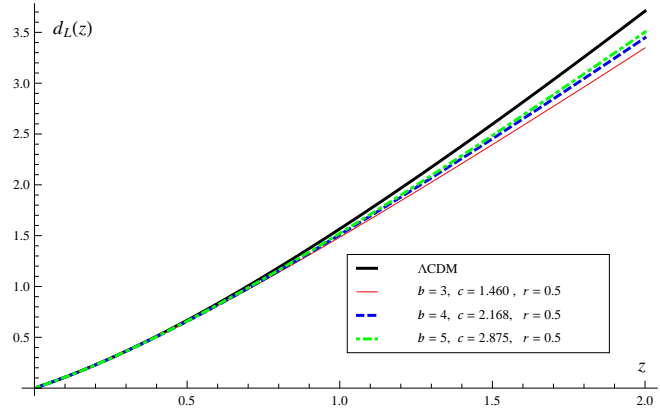}
\caption[Short caption for figure 7]{\label{fig:luminosity0.5} {Luminosity distance for $r=0.5$
and for different values of $b$. It can be observed that for large values of $b$ the curves
approach to $\Lambda$CDM}.}
\end{center}
\end{figure}

\begin{figure}
\begin{center}
\includegraphics[scale=0.7]{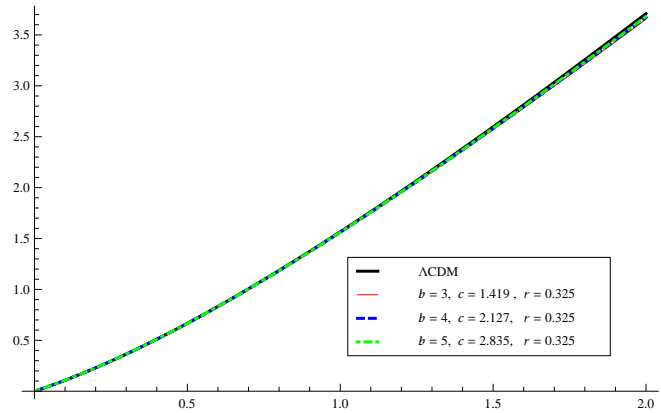}
\caption[Short caption for figure 8]{\label{fig:luminosity0.325} {Luminosity distance for $r=0.325$.
It can be observed that the model curves are very close to $\Lambda$CDM.}}
\end{center}
\end{figure}

The evidence for the late time accelerated expansion of the universe came from Type Ia Supernovae
observations \cite{Riess,Perlmutter} and two important tests to constrain the parameters of
dark energy models in general to the current observations are the luminosity distance and the
Alcock-Paczy\'{n}ski test \cite{AP}.
The luminosity distance is calculated from
\begin{equation}
d_L = (1 + z) \int_0 ^z H_0 \frac{dx}{H(x)},
\end{equation}
where $H(t)= \dot{a}/a$ is the Hubble parameter that satisfies the Friedmann equation
\begin{equation}
\frac{H^{2}}{H^{2}_{0}} = \Omega_m(1 + z)^{3} + \Omega_{(k-essence)},
\end{equation}
where $H_{0}$ represents the Hubble parameter at the current time $t_0$, $\Omega_m$ is the parameter representing the
contribution from matter (dust), and we represent the dark energy component of the universe with the parameter
$\Omega_{(k-essence)} \equiv \rho/\rho_{c}$ where $\rho$ is given by eq. (\ref{k-essence-rho}) and $\rho_{c} = 3H^{2}_{0}/8 \pi G$
is the usual critical density. By means of the the normalization condition $\Omega^{(0)}_{m} + \Omega^{(0)}_{(k-essence)} = 1$ we
can determine the parameter $c$ in terms of the other free parameters of the model. As a result we have $b$ and $r$ as free parameters,
the latter being defined as in Eq. (\ref{quartic-equation}). In figures \ref{fig:luminosity0.5} and \ref{fig:luminosity0.325}
we compare the luminosity distance from our model with the $\Lambda$CDM model.
\begin{figure}
\begin{center}
\includegraphics[scale=0.7]{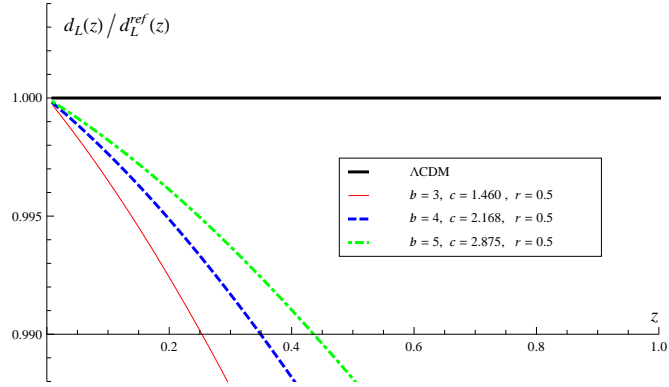}
\caption[short caption for figure 9]{\label{fig:luminosityB0.5}
{$d_L$(KCDM)/$d_L$($\Lambda$CDM) for $r=0.5$ and different values of $b$.
We can observe that the difference between the models is very small}}
\end{center}
\end{figure}
\begin{figure}
\begin{center}
\includegraphics[scale=0.7]{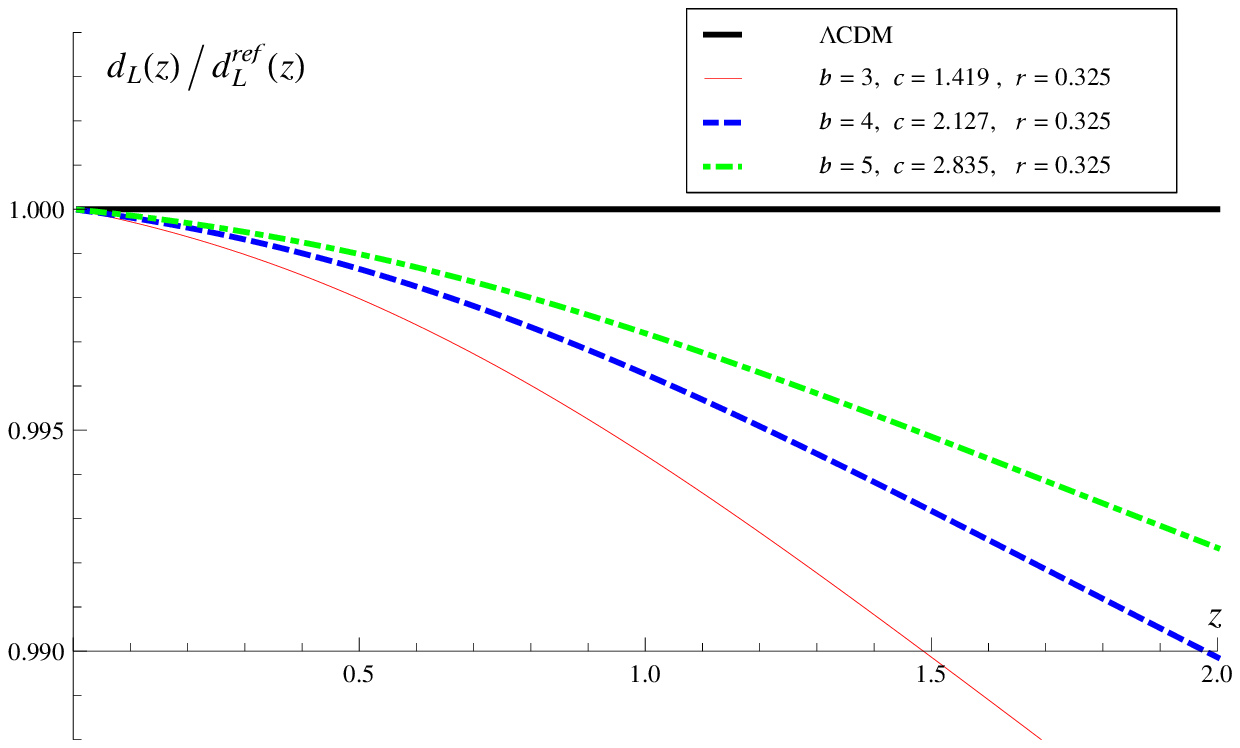}
\caption[short caption for figure 10]{\label{fig:luminosityB0.325}
{$d_L$(KCDM)/$d_L$($\Lambda$CDM) for $r=0.325$ and different values of $b$}}
\end{center}
\end{figure}

From these figures we conclude that the model approaches the $\Lambda$CDM model for large values of $b$ and
for small values of the parameter $r$. A more suitable way for comparing the two models is by means of the quotient
$d_L$(KCDM)/$d_L$($\Lambda$CDM), where KCDM stands for Cold Dark Matter with k-essence as the agent (dark energy)
producing the observed late-time acceleration of the universe.

\begin{figure}
\begin{center}
\includegraphics[scale=0.7]{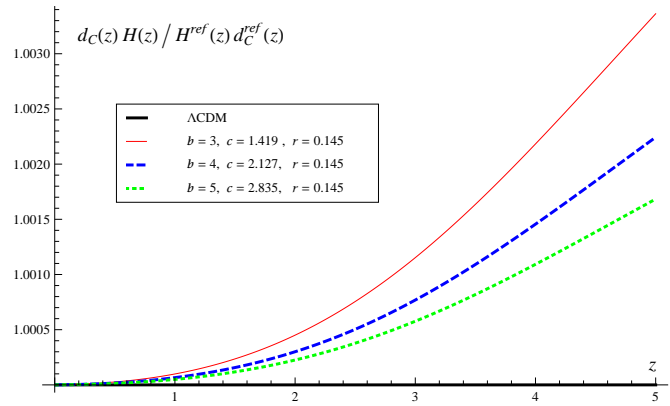}
\caption[short caption for figure 13]{\label{fig:luminosityc0.145} {Alcock-Paczynski test for $r=0.145$. This figure
shows that for larger values of $b$ and small values of $r$ the model curves approaches to the $\Lambda$CDM model.}}
\end{center}
\end{figure}

Some results are shown in figures \ref{fig:luminosityB0.325} and \ref{fig:luminosityB0.5}.
These results confirm clearly the fact that for larger values of $b$ and lower
values for the parameter $r$ the model mimics $\Lambda$CDM.
\begin{figure}
\begin{center}
\includegraphics[scale=0.7]{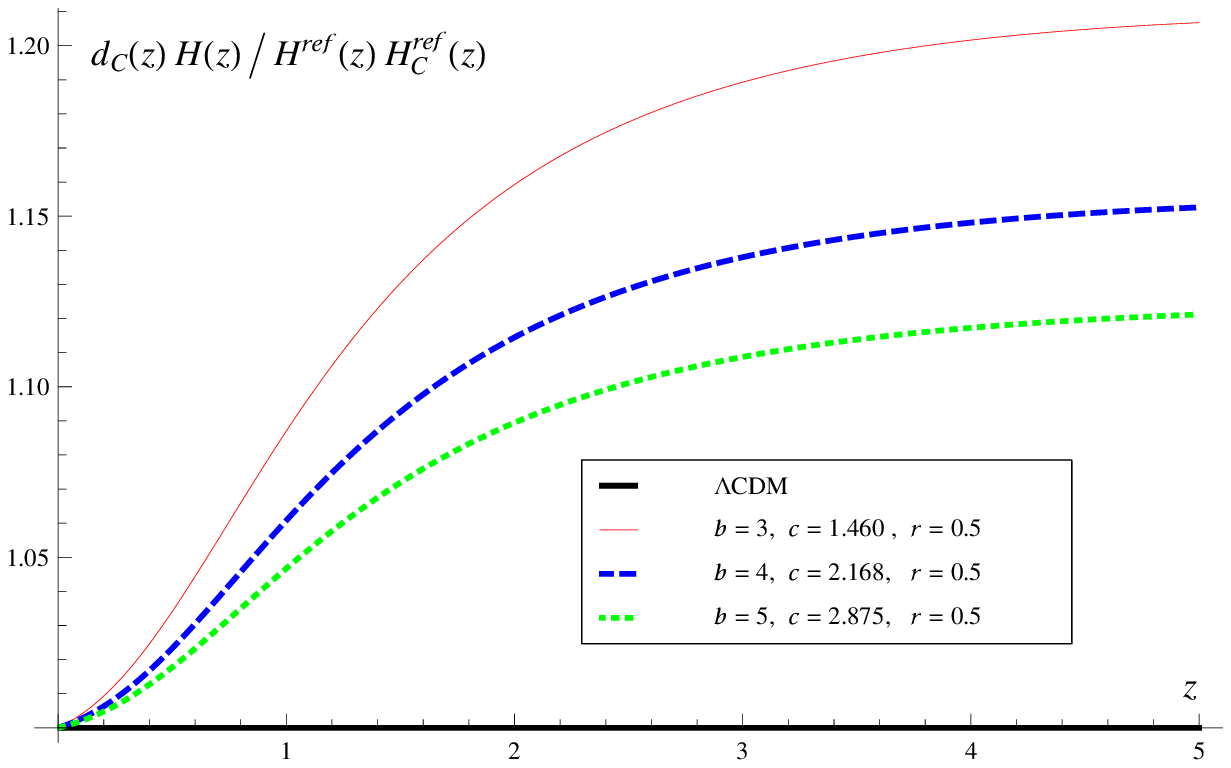}
\caption[short caption for figure 11]{\label{fig:luminosityc0.5} {Alcock-Paczynski test for $r=0.5$}}
\end{center}
\end{figure}
\begin{figure}
\begin{center}
\includegraphics[scale=0.7]{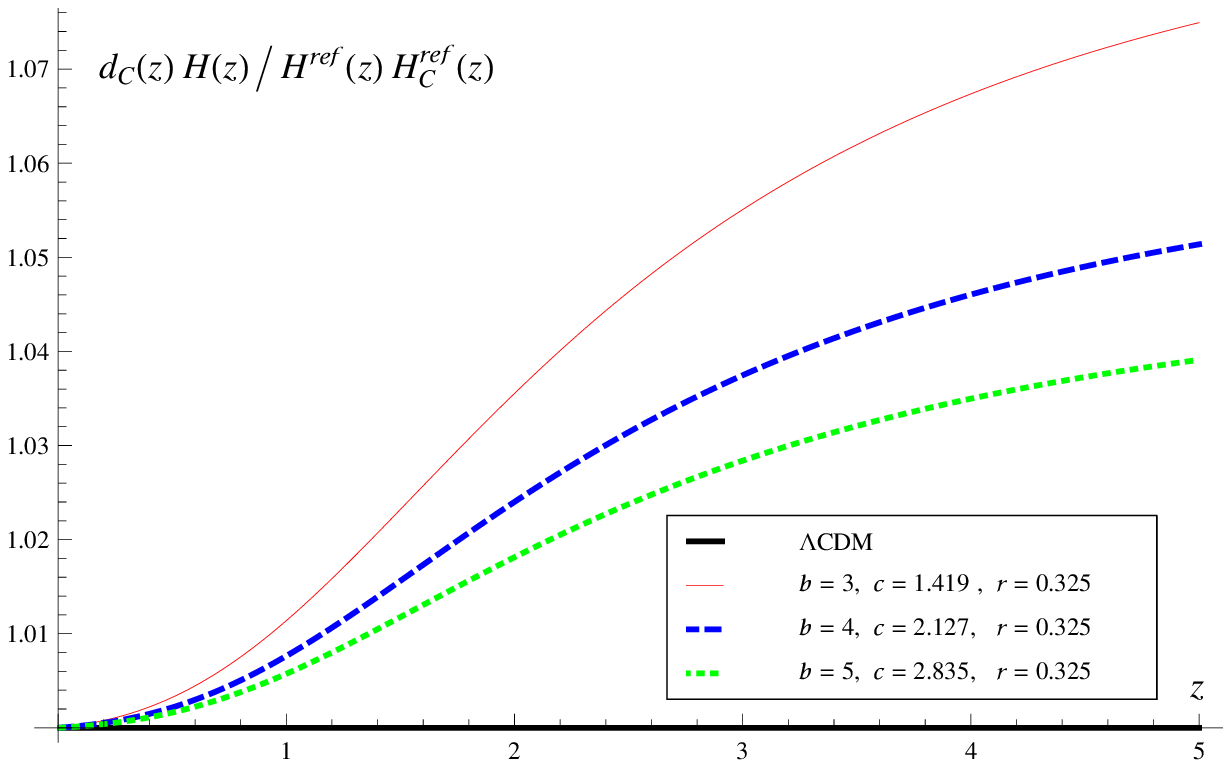}
\caption[short caption for figure 12]{\label{fig:luminosityc0.325} {Alcock-Paczynski test for $r=0.325$}}
\end{center}
\end{figure}
Another useful tool to compare the predictions of our model with standard cosmology
with cosmological constant is the Alcock-Paczy\'{n}ski test \cite{AP}. This test is independent of evolutionary
effects and is a very sensitive estimator for dark energy. Results using this test are presented
in figures \ref{fig:luminosityc0.5}, \ref{fig:luminosityc0.325}, and \ref{fig:luminosityc0.145}.
In this test we can conclude that our model is successful to model dark energy at this level. However another tests
should be applied in order to constraint the values of the free parameters of the model. These tests may allow direct
comparison with supernovae Ia observations and Hubble parameters results. This latter task is under current investigation.

\section{Conclusions}

In this paper we have considered a nonlinear equation of state for a k-essence field inspired
by superconducting topological defects. In this proposal the dark energy content of the universe comes from
a symmetry breaking mechanism of the kind of process that give origin to cosmic topological defects.
We have constrained the range of valid parameters of the model demanding stability
(the nonexistence of imaginary adiabatic speed of sound), the condition of positive
values for the energy density and moreover we have found the regions for superluminal and subluminal
adiabatic speed of sound. It is important to mention that the velocity of the perturbations tends asymptotically to a
constant depending on the parameters used in the model and it can be arbitrarily small.
We have shown the evolution for the equation of state parameter with respect to
the scale factor and, for certain cases, the k-essence field behaves like radiation then it evolves toward
dust and finally it goes asymptotically to cosmological constant. This is an example of a solution with k-essence
that tracks the equation of state of the dominant type of matter. We also considered the evolution of the luminosity
distance $d_{L}$ with respect to redshift, and for this purpose we compare the predictions of our model to those
of standard cosmology with a dark energy content ($\Lambda$CDM). The proposed model has $b$ and $r$ as free
parameters and the general behavior consists in that for large values of $b$ and small ones for $r$ the model is similar
to $\Lambda$CDM model. However there is an specific case when the k-essence field has a positive value for the
equation of state parameter for large values of the scale factor and it does not evolve into a cosmological constant.
Since we consider the evolution of the universe from a $z$-dependent equation of state parameter $w(z)$ the luminosity
distance $d_{L}(z)$ was also compared (normalized) with that for the standard cosmology $\Lambda$CDM using the
Alcock-Paczi\'{n}ski test. As stated before, other tests should be applied in order to constrain the values of
the free parameters of the model. These tests may allow direct comparison with supernovae Ia observations and
Hubble parameters values. This latter task is under a current investigation.
We believe that other kind of superconducting membrane Lagrangians deserve physical exploration within the context of
purely kinetic k-essence models.

\bigskip
{\bf Acknowledgments}

The authors acknowledge partial support from CONACyT under grant J1-60621. R. C. and A. Q. acknowledge support
from COFAA-IPN, EDI and SIP-20161291. E. L. G. acknowledges partial support from PIFI-IPN.
We also thank SNI-M\'exico for partial support.

\bigskip

\vfill\break
\bibliographystyle{unsrt}

\begin{thebibliography}{99}

\bibitem{Riess}
\texttt{Supernova Search Team} collaboration, A.G. Riess et al.,
\emph{Observational evidence from supernovae for an accelerating universe and
a cosmological constant}, \emph{Astron. J.} \textbf{116}  (1998) 1009 [astro-ph/9805201]

\bibitem{Perlmutter}
\texttt{Supernova Cosmology project} collaboration, S. Perlmutter et al.,
\emph{Measurements of $\Omega$ and $\Lambda$ from 42 high redshift supernovae,}
\emph{Astrophys. J.} \textbf{517} (1999) 565 [astro-ph/981233]

\bibitem{Sahni:2004ai}
V. Sahni,
\emph{Dark matter and dark energy}, Lect.\ Notes Phys.\  \textbf{653} (2004) 141
[arXiv:astro-ph/0403324]

\bibitem{Alam:2004jy}
U. Alam, V. Sahni and A. Starobinsky, \emph{The Case for dynamical dark energy revisited},
\emph{JCAP} \textbf{406} (2004) 008 [arXiv:astro-ph/0403687v2]

\bibitem{Sahni:2006pa}
V. Sahni and A. Starobinsky, \emph{Reconstructing Dark Energy},
\emph{Int. J. Mod. Phys.}  \textbf{D} \textbf{15} (2006) 2105 [arXiv:astro-ph/0610026]

\bibitem{Feng:2004ad}
B. Feng, X. L. Wang and X. M. Zhang, \emph{Dark energy constraints from the cosmic age and supernova},
\emph{Phys. Lett.} \textbf{B 607} (2005) 35 [arXiv:astro-ph/0404224]

\bibitem{Durrer:2007re}
R. Durrer and R. Maartens, \emph{Dark Energy and dark gravity},
\emph{Gen. Rel. Grav.} \textbf{40} (2008) 301 [arXiv:0711.0077v2]

\bibitem{Wette}
C. Wetterich, \emph{Models for dark energy}, \emph{Nucl. Phys.} \textbf{B} 302 (1988) 668.

\bibitem{Peebles}
B. Ratra and J. Peebles, \emph{Cosmological consequences of a rolling homogeneous scalar field}, \emph{Phys. Rev.}
\textbf{D 37} (1988) 3406

\bibitem{Copeland}
E. Copeland, M. Sami, S. Tsujikawa, \emph{Dynamics of dark energy},
\emph{Int. J. Mod. Phys.} \textbf{D 15} (2006) 1753-1936 [arXiv:hep-th/0603057]

\bibitem{Linder:2007wa}
E.V.Linder,
\emph{The dynamics of quintessence, The quintessence of dynamics},
\emph{Gen. Rel. Grav.} \textbf{40} (2008) 329 [arXiv:0704.2064v2]

\bibitem{Deffayet}
C. Deffayet, G. R. Dvali and G. Gabadadze, \emph{Accelerated universe from gravity leaking to extra dimensions},
\emph{Phys. Rev.}, \textbf{D 65} (2002) 044023 [arXiv:astro-ph/0105068]

\bibitem{SS}
V. Sahni and Y. Shtanov, \emph{Braneworld models of dark energy}, \emph{JCAP} \textbf{0311}, (2003) 014
[arXiv:astro-ph/0202346v3]

\bibitem{InfADM}
C. Armendariz-Picon, T. Damour and V. Mukhanov, \emph{k-Inflation},
\emph{Phys. Lett. B} \textbf{458} (1999) 209-218 [arXiv:hep-th/9904075]

\bibitem{Armendariz-Mukhanov-Steinhardt}
C. Armendariz-Picon, V. Mukhanov, P. J. Steinhardt, \emph{Dynamical solution to the problem
of a small cosmological constant and late-time acceleration}, \emph{Phys. Rev. Lett.} \textbf{85},
(2000) 4438 [arXiv:astro-ph/0004134]

\bibitem{AMS}
C. Armendariz-Picon, V. Mukhanov and P. J. Steinhardt, \emph{Essentials of k-essence},
\emph{ Phys. Rev.} \textbf{D 63} (2001) 103510 [arXiv:astro-ph/006373]

\bibitem{Melchiorri}
A. Melchiorri, L. Mersini, C. Odman and M. Trodden, \emph{The state of the dark energy equation of state},
\emph{Phys. Rev.} \textbf{D 68} (2003) 043509 [arXiv:astro-ph/0211522]

\bibitem{Chiba1}
T. Chiba, T. Okube and M. Yamaguchi, \emph{Kinetically driven quitaessence},
\emph{Phys. Rev.} \textbf{D 62} (2000) 023511 [arXiv:astro-ph/9912463v2]

\bibitem{Chiba2}
T. Chiba,\emph{ Tracking k-essence}, \emph{Phys. Rev.} \textbf{D  66}, 063514 (2002)
[arXiv:astro-ph/0206298v2]

\bibitem{Chimento1}
L. P. Chimento and A. Feinstein, \emph{Power-law expansion in k-essence cosmology},
\emph{Mod. Phys. Lett}. \textbf{A 19} (2004) 761 [arXiv:astro-ph/0305007v2]

\bibitem{Chimento2004}
L. P. Chimento, \emph{Extended tachyon field, Chaplygin gas and solvable k-essence cosmologies},
\emph{Phys. Rev.} \textbf{D  69} (2004) 123517 [arXiv:astro-ph/0311613]

\bibitem{Amendola-Tsujikawa}
L. Amendola and S. Tsujikawa, \emph{Dark Energy: Theory and Observations}, Cambridge University Press, Cambridge, 2010.

\bibitem{Jaume}
J. Garriga and V. F. Mukhanov, \emph{Perturbations in k-Inflation}
\emph{Phys. Lett. B} \textbf{458} (1999) 219 [arXiv:hep-th/9904176]

\bibitem{Scherrer:2004au}
R. J. Scherrer, \emph{Purely kinetic k-essence as unified dark matter},
\emph{ Phys. Rev. Lett.}  \textbf{93} (2004) 011301 [arXiv:astro-ph/0402316v3]

\bibitem{Linder1}
R. de Putter and E. V. Linder, \emph{Kinetic k-essence and quintessence} \emph{Astropart. Phys.}
\textbf{28} (2007) 263-272 [arXiv:astro-ph/0705.0400v2]

\bibitem{Gao}
X. T. Gao and R. J. Yang, \emph{Geometrical diagnostic for purely kinetic k-essence dark energy}
\emph{Phys. Lett. B}, \textbf{687} (2010) 99 [arXiv:gr-qc/1003.2786v2]

\bibitem{holographic}
N. Cruz, P. F. Gonz\'alez-Diaz, A. Rozas-Fern\'andez and G. S\'anchez,
\emph{Holografic kinetic k-essence model},
\emph{Phys. Lett. B} \textbf{679} (2009) 293-297 [arXiv:gr-qc/0812.4856v2]

\bibitem{crossing}
S. Sur and S. Pas, \emph{Multiple kinetic k-essence, phantom barrier crossing and stability},
\emph{JCAP}, \textbf{0901} (2009) 007 [arXiv:astro-ph/0806.4368v3]

\bibitem{Witten}
E. Witten, \emph{Supercondcuting strings} \emph{Nucl. Phys. B} \textbf{249} (1985) 557.

\bibitem{Alexbook}
A. Vilenkin and E. P. S Shellard, \emph{Cosmic Strings and Other Topological Defects}, Cambridge University Press,
Cambridge, 1994.


\bibitem{CarterRev}
B. Carter, \emph{``Brane dynamics for treatment of cosmic strings and vortons,''} Lectures given at 2nd Mexican School
on Gravitation and Mathematical Physics, Tlaxcala, Mexico, 1-7 Dec 1996 [arXiv:hep-th/9705172]


\bibitem{Peter2}
P. Peter, Superconducting cosmic string: \emph{Equation of state for spacelike and timelike current in the neutral limit},
\emph{Phys. Rev. D}, \textbf{45} (1992) 1091.

\bibitem{supersonic}
B. Carter and P. Peter, \emph{Supersonic string models for Witten vortices}, \emph{Phys. Rev. D} \textbf{52} (1995) R1744
[arXiv:hep-th/9411425]

\bibitem{Lazarides:1985my}
G. Lazarides and Q. Shafi,
\emph{Superconducting membranes}, \emph{Phys. Lett. B} \textbf{159} (1985) 261.

\bibitem{Peter:1995ks}
P. Peter,
\emph{Surface current carrying domain walls},
\emph{J. Phys. A} \textbf{29} (1996) 5125 [arXiv:hep-ph/9503408v1]

\bibitem{Rubakov:1983bb}
V. A. Rubakov and M. E. Shaposhnikov, \emph{Do we live inside a domain wall?}
\emph{Phys. Lett. B } \textbf{125} (1983) 136-138

\bibitem{Nielsen}
H.B. Nielsen and P. Olesen, \emph{Vortex-line models for dual strings}, \emph{ Nucl. Phys.} \textbf{B61}, 45 (1973).

\bibitem{Cordero:2000hx}
R. Cordero and E. Rojas, \emph{Chiral superconducting membranes},
\emph{Int. J. Mod. Phys. A} \textbf{17} (2002) 73 [arXiv:astro-ph/0009140v2]

\bibitem{Gorini-Kamenshchik-Moschella-Pasquier}
V. Gorini, A. Kamenshchik, U. Moschella, and V. Pasquier, \emph{The Chaplygin gas as a model for dark energy},
10$^{th}$ Marcell Grossmann Meeting on Recent Developments in Theoretical and Experimental General Relativity,
Graitation and Relativistic Fiedl Theories. (M6 X MMIII), 20-26 Jul. 2002, Rio de Janeiro, Brazil.
Edit: M. Novello, S. Perez.Bengliaffa, R. Ruffini. pp. 840-859
[arXiv:gr-qc/0403062]

\bibitem{Kamenshchik-Moschella-Pasquier}
A. Kamenshchik, U. Moschella, and V. Pasquier, \emph{An alternative to quintessence}, \emph{Phys. Lett. B}, \textbf{511}, (2001) 265
[arXiv:gr-qc/0103004]

\bibitem{Bento-Bertolani-Sen}
M.C. Bento, O. Bertolani, and A.A. Sen, \emph{Generalized Chaplygin gas, accelerated expansion,
and dark-energy-matter unification} \emph{Phys. Rev. D} \textbf{66}, 043507 (2002)
[arXiv:gr-qc/0202064]

\bibitem{Cordero:1999xd}
R. Cordero and E. Rojas, \emph{Constrained superconducting membranes},
\emph{Phys. Lett. B} \textbf{470} (1999) 45 [arXiv:astro-ph/0009139v1]

\bibitem{superhybrid}
R. Cordero and A. Queijeiro, \emph{Superconducting hybrid extended objects},
\emph{J. Phys. A: Math. Gen.} \textbf{34} (2001) 3393.

\bibitem{Odintsov1}
S. Nojiri and S. Odintsov, \emph{Final state and thermodynamics of dark energy universe}, \emph{Phys. Rev. D},
\textbf{70} (2004) 103522 [arXiv:hep-th/0408170]

\bibitem{Odintsov2}
S. Nojiri, S. Odintsov, S. Tsujikawa, \emph{Properties of singularities in the (phantom) dark energy universe},
 \emph{Phys. Rev. D}, \textbf{71} (2005) 063004 [arXiv:hep-th/0501025]

\bibitem{Odintsov3}
S. Capozziello and S. Nojiri, S. Odintsov, \emph{Unified phantom cosmology: Inflation, dark energy and dark matter
under the same standard}, \emph{Phys. lett. B}, \textbf{632} (2006) 597-604 [arXiv:hep-th/0507182]

\bibitem{Odintsov4}
A. Astashenok, S. Nojiri, S. Odintsov, and A. Yurov, \emph{Phantom cosmology without Big Rip singularity},
\emph{Phys. lett. B}, \textbf{709} (2012) 396-403 [arXiv:gr-qc/1201.4056]

\bibitem{Odintsov5}
S. Nojiri and S. Odintsov, \emph{Inhomogeneous equation of state of the universe: Phantom era,
future singularity, and crossing the phantom barrier}, \emph{Phys. Rev. D}, \textbf{72} (2005) 023003
[arXiv:hep-th/0505215]

\bibitem{Odintsov6}
K. Bamba, S. Capoziello, S. Nojiri, and S. Odintsov, \emph{Dark energy cosmology: the equivalent description via
different theoretical models and cosmography tests.} \emph{Astrophys. Space Scie.} 342,
(2012) 155-228 [arXiv: gr-qc/1205.3421]



\bibitem{superluminal1}
E. Babichev, V. Mukhanov and A. Vikman, \emph{k-Essence, superluminal propagation, causality
and emergent geometry}, JHEP \textbf{0802} (2008) 101 [arXiv:hep-th/0708.056]

\bibitem{Bruneton:2006gf}
J. P. Bruneton,
\emph{On causality and superluminal behavior in classical field theories: Applications to k-essence
theories and MOND-like theories of gravity}, \emph{Phys. Rev. D} \textbf{75} (2007) 085013 [arXiv:gr-qc/0607055v2]

\bibitem{Kang:2007vs}
J. U. Kang, V. Vanchurin, S. Winitzki,
\emph{Attractor scenarios and superluminal signals in k-essence cosmology},
\emph{Phys.Rev. D} \textbf{76} (2007) 083511 [arXiv:gr-qc/0706.3994]

\bibitem{Bonvin:2006vc}
C. Bonvin, C. Caprini and R. Durrer,
\emph{A no-go theorem for k-essence dark energy},
\emph{Phys. Rev. Lett.} \textbf{97} (2006) 081303 [arXiv:astro-ph/0606584]

\bibitem{Gorini:2007ta}
V. Gorini, A. Y. Kamenshchik, U. Moschella, O. F. Piattella, A. A. Starobinsky,
\emph{Gauge-invariant analysis of perturbations in Chaplygin gas unified models of dark matter and dark energy},
\emph{JCAP} \textbf{0802} (2008) 016 [arXiv:astro-ph/0711.4242]

\bibitem{AP} C. Alcock, and B. Paczy\'{n}ski, \emph{ An evolution free test for non-zero cosmological constant},
\emph{Nature}, \textbf{281}, (1979) 358.

\end{thebibliography}

\end{document}